\documentclass[12pt]{iopart}
\usepackage{iopams, amssymb,graphicx,bbm}
\usepackage{xcolor}
\usepackage{ulem}

\begin{document}

\title[Phase modulated two-photon Fourier transform spectroscopy]{Two-photon excitation spectroscopy of 1,5--Diphenyl-1,3,5-hexatriene using phase modulation }

\author{Pushpendra Kumar$^{1}$, Khadga Jung Karki$^{1,*}$}

\address{$^1$Chemical Physics, Lund University, Naturvetarv\"agen 16, 22362 Lund, Sweden}
\ead{$^*$Khadga.Karki@chemphys.lu.se}
\vspace{10pt}
\begin{indented}
\item[]\date{}
\end{indented}

\begin{abstract}
  We have used two-photon Fourier transform spectroscopy to investigate the first singlet excited state (S1) of a prototypical polyene molecule 1,5 -- Diphenyl-1,3,5-hexatriene. As the S1 state in the polyenes is a one-photon forbidden transition, its structure of the vibrational levels cannot be studied using resonant linear excitation. Although this level is accessible with two-photon excitation, previous studies done by using wavelength tunable pulsed lasers did not have enough resolution to investigate the details of the vibrational levels. In Fourier transform spectroscopy, one uses a pair of laser beams to excite the sample. The measurements are done by varying the time delay between the pulses. The spectral resolution is given by the inverse of the maximum time delay rather than the spectral width of the pulses. We have used the method to investigate  the vibrational levels of the S1 state.  In our implementation, we have used phase modulation technique to carry out the measurements in the rotating frame, which require less data points along the time delay thereby significantly reducing the measurement time.
\end{abstract}

%
\vspace{2pc}

\section{Introduction}

Two-photon absorption (TPA) was theoretically predicted by Maria G\"oppert-Mayer in 1931,\cite{MAYERS_1931} and observed experimentally in the early 1960’s.\cite{GARRET_1961} Since the development of the pulsed lasers, TPA has been used in a variety of photonic and biological applications such as two-photon laser scanning microscopy,\cite{WEBB_1990} upconversion lasing,\cite{SUN_2010,PAGANI_1997} optical power limiting,\cite{PERRY_1997} 3-D microfabrication,\cite{PARK_2008,DUAN_2007} ultrafast pulse characterization,\cite{SAUERBREY_1997} optical data storage,\cite{WEBB_1991,SPAHNI_2007} sensors\cite{PERRY_2004} and photodynamic therapy.\cite{PRASAD_2007} TPA has also been important in spectroscopy of atoms, molecules and larger systems. Some unique features of TPA include Doppler-free spectroscopy\cite{WIEMAN_1975} and spectroscopy of electronic states that are not directly accessible by one photon transition.\cite{BONIN_1984}

Different methods have been used to measure the TP-excitation spectra. In the early days, tunable narrow band high power lasers were used.\cite{MARTIN_1974} This technique is deprecated as many different types of nonlinear interactions, apart from TPA, can contribute because of the long pulse duration. It is now accepted that  impulsive excitation by ultrashort pulses that are  substantially shorter than a picosecond, yield accurate results. However, when the ultrashort pulses are used for spectroscopy by scanning the wavelength, spectral resolution is limited by the laser bandwidth. Two-photon Fourier transform (FT) spectroscopy has been implemented to circumvent this limitation.\cite{JOFFRE_2004} Similar methods have also been combined with IR spectroscopy to obtain detailed information about the vibrational levels in molecular systems.\cite{TOKMAKOFF_2016,TOKMAKOFF_2018} 

In the FT spectroscopy, one uses two broadband beams (usually ultrashort pulses), and records the response of the system as a function of time delay between the beams. The Fourier transforms of the transients give the spectra in the frequency domain. The spectral resolution is given by the inverse of the maximum time delay, which can easily be longer than the dephasing time of the system. Nevertheless, the technique is difficult to implement in the visible and ultraviolet wavelength regions mainly because it is an interferometric method that demands an extremely stable setup. The data have to be acquired at very short intervals of the time delay to follow the  dense interference patterns. Here, we use phase modulation to alleviate these difficulties.\cite{MARCUS2006,STIENKEMEIER_2015,STEINKEMEIER_2015B}   More recently, dual frequency combs have also been used in high resolution two-photon spectroscopy.\cite{PIQUE_2014} 

Phase modulation has been used previously in the FT spectroscopy of excited states under resonant one-\cite{MARCUS2006} or sequential multi-photon excitations.\cite{STIENKEMEIER_2015,STEINKEMEIER_2015B} The method uses two beams whose phases are coherently modulated at slightly different radio frequencies, here denoted as $\phi_1$ and $\phi_2$. The linear response of a system excited by such beams show modulation at the frequency $\phi_{21}=\phi_2-\phi_1$. Depending on the order of the interaction, the nonlinear responses show modulations at $n\phi_{21}$, where $n$ is an integer.\cite{STEINKEMEIER_2015B,KARKI_2016A,KARKI_2016C,KARKI_2018B} The phases and the amplitudes of the modulated signals are measured with respect to a reference whose phase evolves at the optical frequency close to that of the excitation beams. If the state contributing to the response of the system is resonantly excited at the frequency $\omega$  and the reference is at $\omega_{ref}$, the time domain interferogram of the modulated signal at $\phi_{21}$ evolves at the frequency $\omega-\omega_{ref}$. As the period of this interferogram is substantially longer than that of the optical frequency itself, it can be easily recorded by varying the time delay between the beams in a normal setup. This eliminates the complexity of stabilizing the setup to an interferometric precision at the optical frequencies. Moreover, the slowly evolving interferogram can be traced with substantially less measurement points along the time delay.\cite{MARCUS2006,STIENKEMEIER_2015}   Although, phase modulation techniques are well established under resonant excitation conditions,\cite{MARCUS2006,MARCUS2007,MARCUS_2012,MARCUS_2013,KARKI_2014C} its implementation under TP-excitation requires extra consideration about the different modulated signals.  Out of the two modulated responses that are observed at the frequencies $\phi_{21}$ and $2\phi_{21}$, only the later contains spectroscopic information about the system. In this case, the reference evolves at twice the frequency of the reference used in the measurements of the linear response (see section below for the details). Our results contradict previous reports\cite{STEINKEMEIER_2015B}  where both the signals have been used to measure the decay of coherence between the ground state and the final excited state in rubidium.

\begin{figure}[htbp]
	\centering
	\includegraphics[width=8cm]{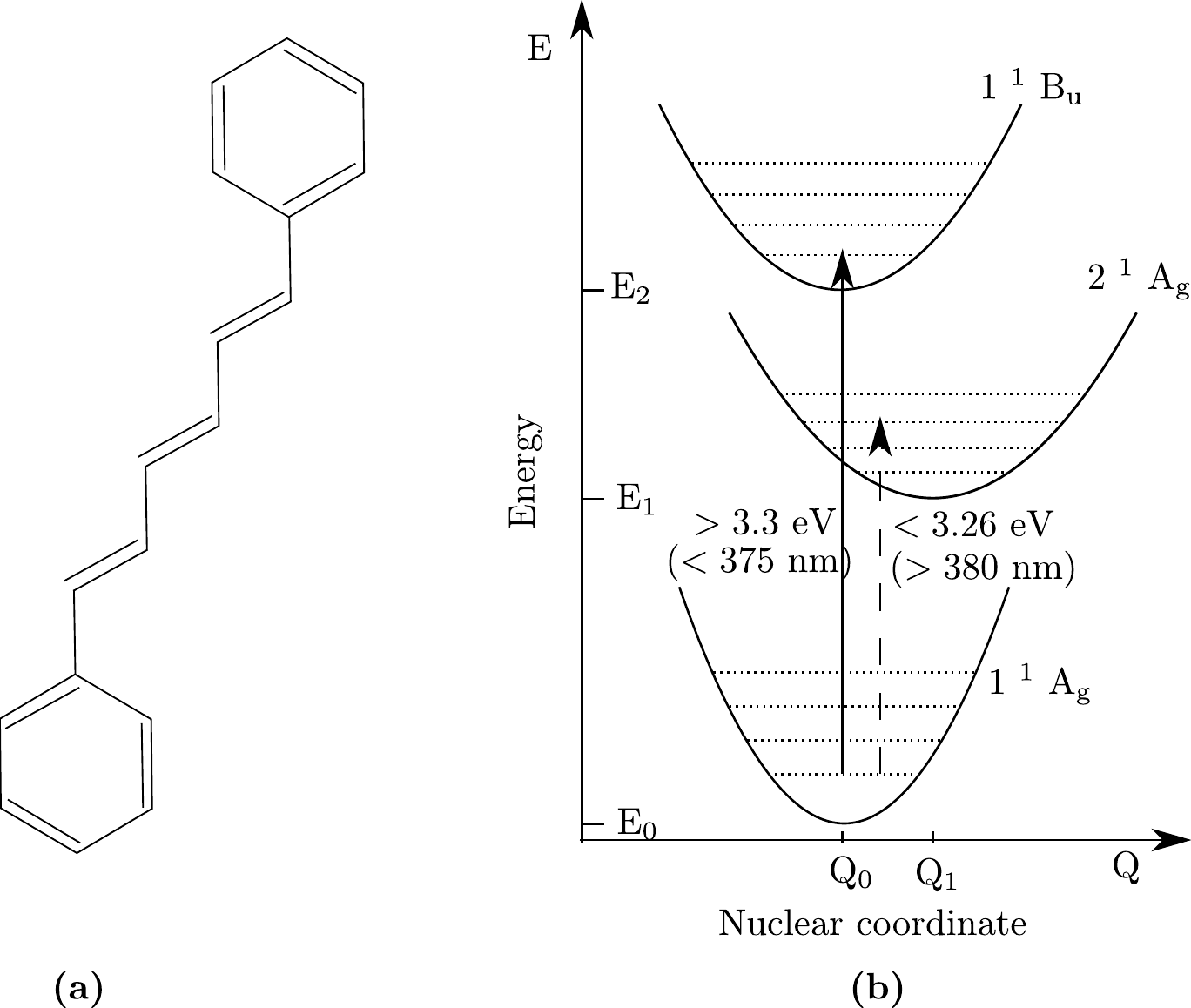}
	\caption{(a) Molecular structure and (b) schematic of the electronic structure of DPH based on the commonly accepted model ( see references \cite{KOHLER_1972,KARPLUS_1972,SPIGLANIN_1984}).}
	\label{FIG0}
\end{figure}

We have used phase modulated TP-excitation to measure the excited spectra of  1,5--Diphenyl-1,3,5-hexatriene (DPH) at the wavelengths between 370 nm to 410 nm. DPH is a member of $\pi$-conjugated $\alpha,\omega$-diphenylpolyenes. Spectral properties of the diphenylpolyenes have been studied since 1930.\cite{SEITZ_1935} The fluorescence spectrum of the diphenylpolyenes have large Stokes shift with respect to the first absorption peak. The strong absorption peak in one-photon process is due to the 1$A_g$ (S$_0$)$\rightarrow 1B_u$ (S$_2$) transition, where the ground state is 1$A_g$. Based on the weak transitions observed at lower energies,\cite{KOHLER_1972} it was proposed that, for the diphenylpolyenes with a long chain length, the 2$A_g$ (S$_1$) state lies below the 1$B_u$.\cite{KARPLUS_1972} The Stokes shift is explained by assigning the emission to the slow $2A_g\rightarrow 1A_g$ transition (see Fig.\ref{FIG0} for the schematic of the energy levels in DPH).\cite{KOHLER_1987,KOHLER_1988,ITOH_2009,YURIKO_2011}  An alternative model has also been proposed in which the emission is assumed to occur via the relaxed structure of 1$B_u$ state.\cite{CATALAN_2003,CATALAN_2017} Irrespective of the model, TP-excitation spectra acquired by scanning the frequency of the laser has shown strong TPA at energies below the 1$B_u$ state.\cite{LEROI_1978} 
Our results, in general, agree with the previous findings. In addition, higher spectral resolution in our measurements show contributions from a number of vibrational levels in the TP-excitation spectra. 

\section{Experimental setup}
\begin{figure}[htbp]
\centering
\includegraphics[width=13cm]{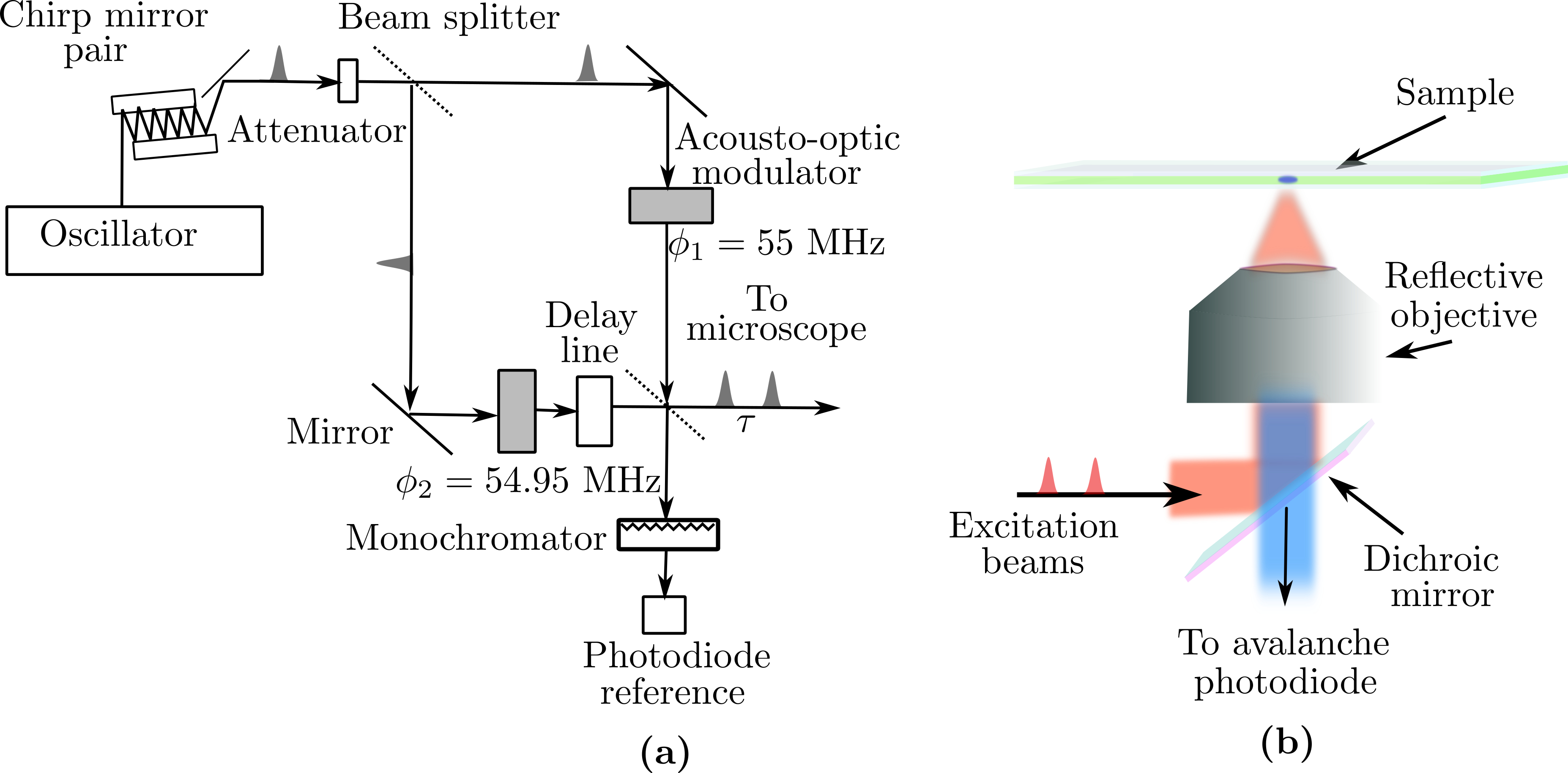}
\caption{(a) Schematics of the optical setup and (b) the microscope setup.}
\label{FIG1}
\end{figure}

The experimental setup is shown in Fig.\ref{FIG1}(a), and is described in detail elsewhere.~\cite{KARKI_2016A,KARKI_2016C,KARKI_2017B} In short, a chirp precompensated pulsed laser beam (wavelength centered at 780 nm with a FWHM of 50 nm and repetition rate of about 73 MHz) is split into two beams, each of which pass through one of the arms of a Mach-Zehnder interferometer. Acousto-optic modulators (AOMs) placed on each of the arms of the interferometer modulate the phases of the beams at the frequencies $\phi_1=54.95$ MHz and $\phi_2=55$ MHz, respectively. The modulations of the phases lead to the shift in the frequencies of the beams: $\omega \rightarrow \omega+\phi_1=\omega_1$ and $\omega \rightarrow \omega+\phi_2=\omega_2$. As the optical frequencies are in the range of few hundreds of THz, the slight shifts cannot be observed using a spectrometer. However, when the two beams are combined using a beam splitter the resulting intensity modulates at the difference frequency $\phi_{21}=\phi_2-\phi_1=50$ kHz. One of the outputs of the beam splitter is directed to an inverted microscope, and focused onto the sample ( 100 $\mu$mol DPH dissolved in toluene) by using a reflective objective (X36, NA=0.5). The two-photon induced photoluminescence (PL) is collected by the same objective and separated from the excitation beams by a dichroic mirror (short pass with cutoff at 670 nm). The PL is detected by an avalanche photodiode (APD). Another short pass filter (cutoff at 600 nm) is placed before the APD to further suppress the stray light reaching the detector. The second output of the beam splitter is sent to a monochromator, which spatially disperses the different wavelengths. A narrow section of the spectrum centered at 808 nm is selected using a slit and directed to a photodiode (PD). The signal from the APD and the PD are digitized simultaneously at the rate of 10 MSa/s (mega samples per second).  The phase and the amplitude of the signal from the APD (two-photon PL) is analyzed with respect to the signal from the PD (reference) by using the algorithms of the generalized lock-in amplifier(GLIA).\cite{KARKI_2013A,KARKI_2013C,KARKI_2014B} The two-photon excitation spectra are recorded as a function of the time delay between the two beams. The time domain data are Fourier transformed, and  shifted by the optical frequency of the reference to obtain the spectra in the frequency domain.   

\section{Theory}
\begin{figure}[htbp]
	\centering
	\includegraphics[width=7cm]{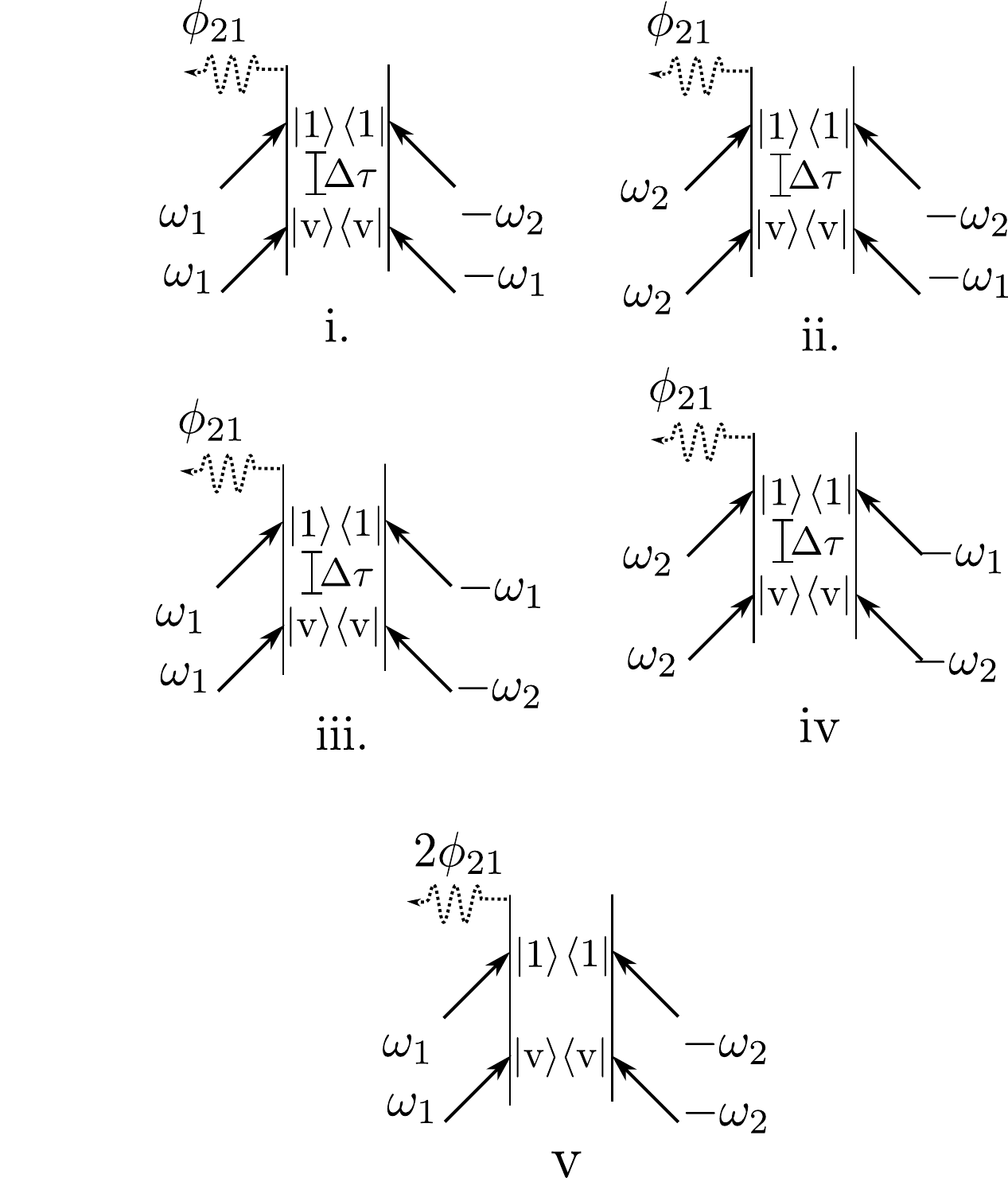}
	\caption{Double sided Feynman diagrams of the different Liouville pathways involved in two-photon excitation. In (i--iv), three field interactions are derived from the same beam and one from the other, while in (v), both the beams interact twice. Interactions (i--iv) and (v) lead to the modulations in the PL at the frequencies $\phi_{21}$ and $2\phi_{21}$, respectively. Note that the interactions that lead to the signal modulation at $\phi_{21}$ need to happen during the pulse overlap indicated as $\Delta \tau$.}
	\label{FIG2}
\end{figure}

It has been shown that the TP-PL induced by a pair of phase modulate beams shows modulations at $\phi_{21}$ and $2\phi_{21}$, where $\phi_{21}$ is the difference in the phase modulation frequency.\cite{KARKI_2016A,KARKI_2016C} The two modulations arise due to the difference in the Liouville pathways, shown in Fig. \ref{FIG2}, that can excite the system to a stationary state. As shown in the figure, four interactions with the field are necessary for a two-photon excitation. In four of the pathways (i--iv), the system interacts three times with one of the beams and once with the other. These pathways lead to the modulation of the PL at the frequency $\phi_{21}$. In the pathway depicted in (v), the system interacts twice with each of the beams. This pathway leads to the modulation of the PL at $2\phi_{21}$. As the transition probabilities in all the pathways are the same, the number of pathways determine the amplitudes of the signals at the two modulation frequencies. Consequently, the two signals have a well defined ratio of $S(\phi_{21}):S(2\phi_{21}) = 4:1$.

  As the diagrams in Fig. \ref{FIG2} show, the signal at the frequency $\phi_{21}$ is present only during the overlap of the two beams, either because the system is in coherence between the excited state and the virtual state after the interaction with one of the beams (pathways i and iv), or because of the time ordering (pathways ii and iii). Thus, this signal does not depend on the dephasing of the  stationary states in the system. Bruder et al. have reported a long lived signal at the modulation frequency $\phi_{21}$ in a multi-photon excitation of rubidium atoms.\cite{STEINKEMEIER_2015B} We emphasize that such signals are only feasible in systems where the intermediate state is an eigen state such that the excitation of the final state occurs by a sequential absorption of multiple photons. In the pathway (v) that gives the signal at $2\phi_{21}$, the system is in coherence between the stationary excited state and the ground state after the interaction with the first beam. As the signal traces the decay of coherence between the excited state(s) and the ground state, it gives excitation spectra of the excited state(s). It has been shown that, for excitation with the unchirped pulses, the excitation spectra $\mathcal{S}(\omega)$, where $\omega=\omega_1+\omega_2$, is given by:\cite{DANTUS_2002}
  
\begin{equation} \label{EQ_spectra}
\mathcal{S}(\omega) = \frac{\textrm{Re}[\mathcal{C}_2(\omega)]}{|A(\omega)|^2},
\end{equation}   
where Re[$\mathcal{C}_2(\omega)$] is the real part of the Fourier transform of the measured coherence decay and $A(\omega)$ represents the second-order electric field amplitude given by 

\begin{equation}\label{EQ2}
A(\omega) \propto \int_{-\infty}^{\infty}|E(\omega)-\Omega| |E(\Omega)| \exp[i\{\phi(\Omega)+\phi(\omega-\Omega)\}]d\Omega,
\end{equation}
where $|E(\Omega)|$ is the laser spectral amplitude and $\phi(\Omega)$ is the phase. The phase contribution can be neglected for unchirped pulses. The spectrum of the second harmonic generated from a thin nonlinear crystal gives the $A(\omega)$. For the excitation at 800 nm, a 10 $\mu$m thick BBO crystal ($\theta=29.2$) can be used as the nonlinear crystals. Previously, the Fourier transform of two-photon photocurrent response from a nonlinear photodiode has also been used to approximate the second-order electric field amplitude.\cite{JOFFRE_2004} However, this method is only suitable if the dephasing times of excited states in the molecules are substantially longer than the dephasing time in the electronic states of the photodiode. Note that if the pulses are not chirped then the second-order electric field amplitude can be easily approximated by using Eq.\ref{EQ2}.   


\section{Results and discussion}

\begin{figure}[htbp]
\centering
\includegraphics[width=15cm]{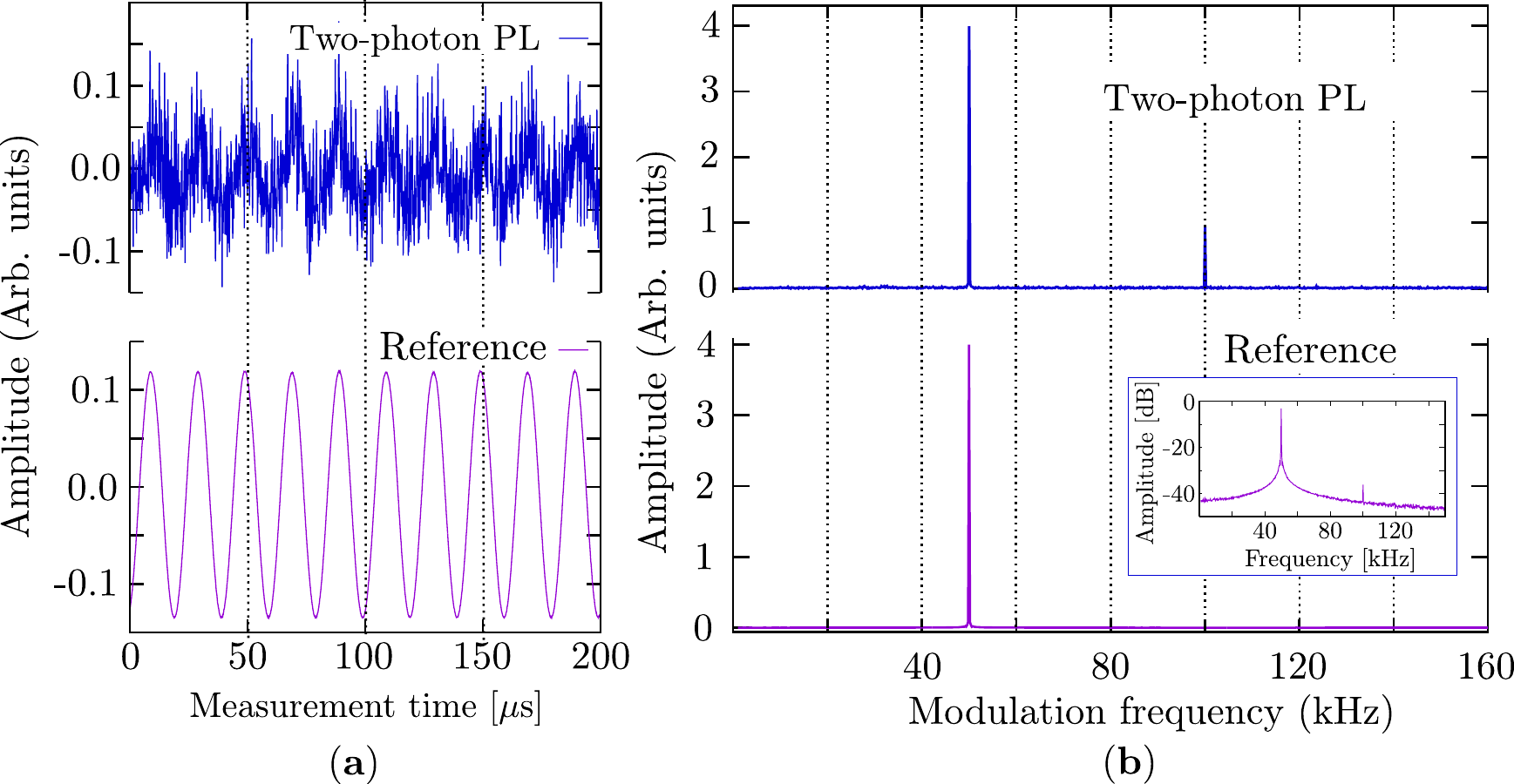}
\caption{Sample of the raw data from TP-excited photoluminescence from DPH (top left) and reference signal from the photodiode (bottom left). The corresponding FFTs are shown on the right.}
\label{FIG_FND_2P2A}
\end{figure}

Fig. \ref{FIG_FND_2P2A} (\textbf{a}) shows the TP-excited photoluminescence signal from the solution of DPH (top) and the response of the photodiode to the reference beam (bottom) measured at the overlap between the pulses. Both the traces show clear modulations. The corresponding Fourier transforms on the right (\textbf{b}) show that the emission from the DPH solution has modulations at two frequencies, 50 kHz and 100 kHz. The ratio of the amplitudes of the two signals is 4:1, which agrees with the pathways shown in Fig. \ref{FIG2}. The reference shows a major peak at 50 kHz from the linear interaction with the beams, and a minor peak at 100 kHz ( visible only in the logscale shown in the inset) due to nonlinear interactions. The peak at 100 kHz is about four orders of magnitude smaller than the peak at 50 kHz.    

The amplitude and the phase of the emission signals at 50 and 100 kHz can be directly extracted from the Fourier transforms (usually FFT) shown in Fig. \ref{FIG_FND_2P2A})(\textbf{b}), or by using the (generalized) lock-in detection techniques. In all the methods, the phases of the signals are measured with respect to the reference. This is straight forward when analyzing the signal at 50 kHz. For the signal at 100 kHz, the phase of the reference at the corresponding frequency can be constructed in different ways. In the traditional lock-in detection, one generates a second harmonic of the photodiode signal at 50 kHz by using a phase locked loop, and uses it as the reference. When using the FFTs or the GLIA, other possibilities such as subtracting twice the phase of the reference at 50 kHz from the signal at 100 kHz, or using the weak reference at 100 kHz can be used. The modulation at 100 kHz is usually present if the reference beam is strong enough to induce nonlinear response in the photodiode. We have used the latter method in our analysis.     

\begin{figure}[htbp]
	\centering
	\includegraphics[width=15cm]{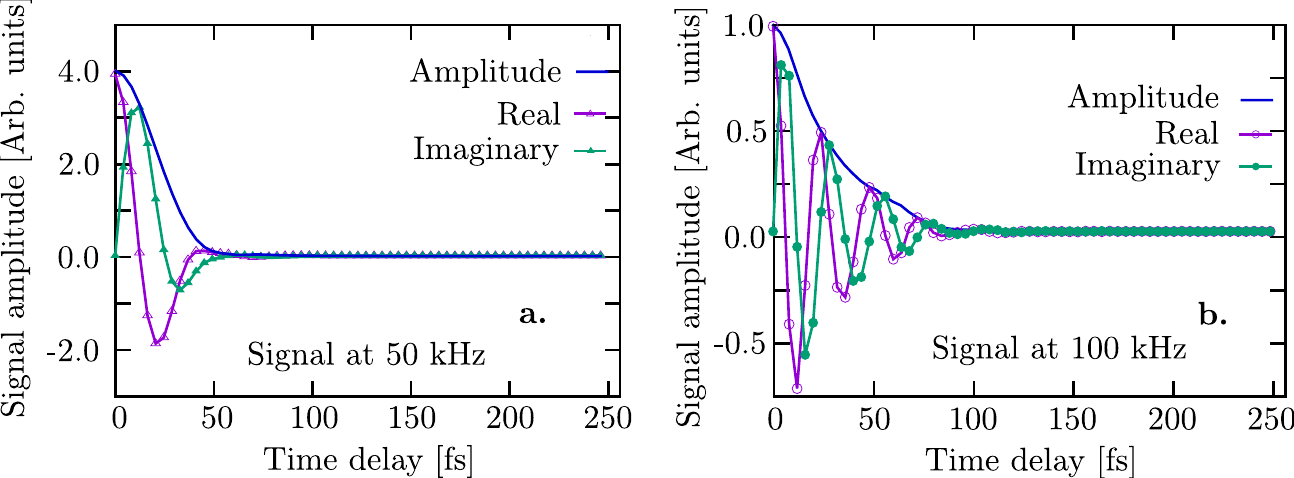}
	\caption{Real and imaginary parts of the TP-photoluminescence signals at the modulation frequencies $\phi_{21}=50$ kHz and $2\phi_{21}=100$ kHz.}
	\label{FIG3}
\end{figure}

The real and imaginary parts of the interferograms of the signals at 50 and 100 kHz are shown in Fig. \ref{FIG3}. The signal at 50 kHz decays quickly within 50 fs as this signal is present only during the overlap of the pulses. 
In constrast, the signal at 100 kHz, shown in Fig.\ref{FIG3}(\textbf{b}), persists even beyond 100 fs of the time delay. The longer decay time clearly indicates that the signal at 100 kHz includes the coherent response from the sample induced by the excitation of the eigenstates. 

It is customary to adjust the phase of the interferograms such that the imaginary part of the signals at the zero time delay is zero\cite{MARCUS2006} (as shown in Fig.\ref{FIG3}). This also maximizes the real part of the signal at the zero time delay. The two-photon absorption spectra in the frequency domain is obtained by the Fourier transform of the time domain signal given in Fig.\ref{FIG3} (\textbf{b}). As the consequence of the phase adjustment of the time domain signal, the imaginary part of the signal in the frequency domain gives the dispersive component and the real part gives the absorptive component of the two-photon excitation spectra. Here, we analyze only the absorptive part of the spectrum in order to compare with the previous measurements.

\begin{figure}[htbp]
\centering
\includegraphics[width=8cm]{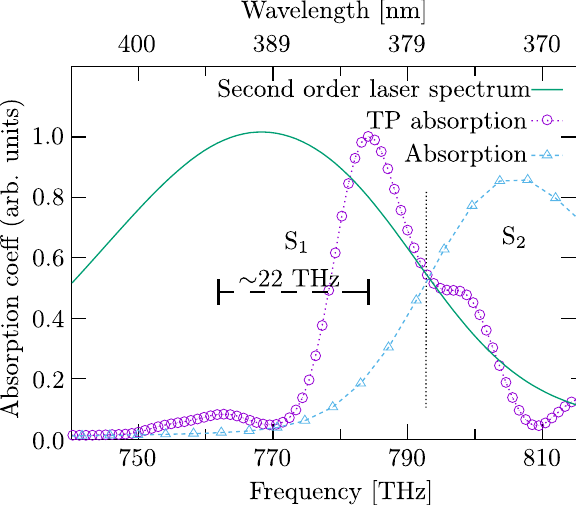}
\caption{ Fourier transform two-photon excitation spectrum using phase modulation (circles and dotted line),  linear absorption spectrum (triangles and dashed line) of DPH, and second order laser spectrum (solid line). The two-photon excitation spectrum, which has been normalized by the second-order laser spectrum, shows prominent peaks at energies lower than the single photon excitation peak. }
\label{FIG5}
\end{figure}

The two-photon excitation spectrum obtained from the Fourier transform of the signal at 100 kHz normalized by the second order laser spectrum is shown as the circled points and the dotted line in Fig.\ref{FIG5}. The spectrum has been shifted by 2$\omega_{\textrm{ref}}$ in order to compare with the one photon absorption spectrum. The one photon absorption spectrum is shown by  the triangles and the dashed line.The second order spectrum is presented by the solid line. The first peak of the one photon absorption, which corresponds to the transition from the ground state S$_0$ to the first vibrational level in the second excited state S$_2$, is at 373 nm (804 THz). The major peak in the TP-excitation spectrum is at 382 nm, which clearly is at energy below the manifold of the S$_2$ state. The previous report of TP-absorption spectra of DPH done by a direct scanning of the excitation wavelength also showed a peak at energies below the one photon transition, which was assigned to the transition to the S$_1$ state. S$_0 \rightarrow$ S$_1$ transition by one photon absorption is assumed to be forbidden due to symmetry. However, as the previous measurements were done using a light source with rather long pulses (few nanoseconds), other processes such as the population of the higher vibrational levels in the ground state and the subsequent excitation to the higher electronic levels during the pulse duration could also contribute to the apparent spectral features below the lowest transition observed in the linear absorption spectra. In our measurements, the pulse duration or more appropriately the autocorrelation can be approximated from the signal at 50 kHz. As the coherence decay observed in the signal at 100 kHz in Fig. \ref{FIG3} (b) occurs at longer time scale, the effects of the pulse duration on the spectral features can be neglected in our measurements. Clearly, our results support the previous reports. Moreover, the TP-excitation spectra in Fig.\ref{FIG5} show multiple peaks below the one photon transition.  Among them, two peaks at  763 THz and 785 THz are separated by 22 THz (730 cm$^{-1}$). These peaks can be compared with the vibrational modes in the ground state of DPH.\cite{AROCA_2007} Importantly, the ground state vibrational modes around 757 cm$^{-1}$ are close to the separation of the peaks we have observed in our spectra. Note that as the widths of peaks are about 13 THz (430 cm$^{-1}$) most of the modes with small amplitudes in between 500 to 2000 cm$^{-1}$ are not clearly visible in our measurements. The broad linewidths is due to the fast dephasing of the coherence at the room temperature. Measurements at low temperatures may be necessary to resolve such modes. The difference of 27 cm$^{-1}$ between the dominant peaks suggests that the corresponding potential energy surface of the S$_1$ state has smaller curvature compared to the ground state.  Miki et al. have investigated vibronic coupling in the excited states of analogous carotenoids.\cite{MOTZKUS_2016} They have observed similar downshift in the frequency of the vibrational modes in the S$_1$ state when compared to the modes in the S$_2$ for the carotenoids with short chain length. Modes at higher frequencies lie beyond the bandwidth of the laser.  

Here, we emphasize that the linewidths in the TP-excitation spectra measured by the FT spectroscopy is not limited by the spectral resolution of the instrument. The resolution in the FT spectroscopy is given by the maximum time delay between the beams in the measurement. Nevertheless, the direct method that was used previously\cite{JOFFRE_2004} can be constrained by the actual measurement time. In the direct method, the sampling should be dense enough to follow the coherent oscillations at the frequencies of the TP transitions. This means for a typical molecule with excited states above 3 eV, the sampling should be done at an interval of 400 as, which may not be feasible if the coherence is long lived. Phase modulation allows one to systematically undersample the coherence decay by appropriately choosing the reference wavelength. As, shown in Fig.\ref{FIG3} (b), measurements sampled at the interval of 4 fs can yield the desired TP-excitation spectra. The undersampling may be crucial in the measurement of FT TP-excitation spectra of atoms and small molecules, where the linewidths are narrower and the corresponding coherences are long lived.\cite{WIEMAN_1975}       

\section{Conclusions}
To summarize, we have shown that phase modulation can be used in conjunction with fluorescence detected TP-FT spectroscopy in order to systematically reduce the measurement points along the coherence time. Out of the two signals that can be observed at the modulation frequencies, $\phi_{21}$ and $2\phi_{21}$, the measurement of the signal at $2\phi_{21}$ as a function of the time delay between the two excitation pulses exhibits coherences between the ground and the excited states. Thus the Fourier transform of the time domain signal provides the TP-excitation spectra. We have used the method  to investigate the TP-excitation spectra of DPH. Although our results broadly agree with the previous reports and show significant TP-excitation at energies below the one-photon transition, the spectra obtained from the new method shows multiple peaks indicative of transitions to different vibrational levels. In general, the phase modulation in FT TP-excitation can enable high resolution spectroscopy of atoms and molecules within reasonable measurement time.  

\textbf{Acknowledgements.}~~
Financial support from the Swedish Research Council (VR), the Crafoord Foundation and Lund Laser Center is gratefully acknowledged.

\section{References}
\providecommand{\newblock}{}

\end{document}